# PDRS: A Fast Non-iterative Scheme for Massive Grant-free Access in Massive MIMO

Yihua Ma, Zhifeng Yuan, Weimin Li, and Zhigang Li

*Abstract*—Grant-free multiple-input multiple-output (MIMO) usually employs non-orthogonal pilots for joint user detection and channel estimation. However, existing methods are too complex for massive grant-free access in massive MIMO. This letter proposes pilot detection reference signal (PDRS) to greatly reduce the complexity. In PDRS scheme, no iteration is required. Direct weight estimation is also proposed to calculate combining weights without channel estimation. After combining, PDRS recovery errors are used to decide the pilot activity. The simulation results show that the proposed grant-free scheme performs good with a complexity reduced by orders of magnitude.

*Index Terms*—Grant-free, massive MIMO, mMTC, pilot detection reference signal, direct weight estimation

## I. Introduction

In massive machine type communications (mMTC) [1], massive users sporadically transmit short packets, which makes the scheduling overheads outweigh the actual data payloads. Grant-free access [2]-[7] is used to reduce overheads and increase the spectral efficiency. To support massive access, massive multiple-input multiple-output (MIMO) is promising as it provides a high spatial degree of freedom [2]-[4].

Classical orthogonal pilot provides only a limited pool size, and non-orthogonal pilot is usually used in grant-free transmissions. There are two major kinds of grant-free massive MIMO schemes: (a) the pre-configuration [2]-[5] scheme where each user is allocated with an identical non-orthogonal pilot, and (b) the random selection scheme where each user randomly selects the pilot [6] [7]. Obviously, the random selection scheme requires the pilot pool to be very large to make pilot collision probability acceptable. Special non-orthogonal pilot design named multi-pilot [7] was used to greatly expand the pilot pool. In this letter, the pre-configuration scheme and generalized non-orthogonal pilot design are considered.

In pre-configuration scheme, the pilot of each user is identical, and the base station obtains both user indices and channel information via pilots. Compressed sensing (CS) [8] methods are usually used to do joint user detection and channel estimation. It is a multiple-measurement vector (MMV) problem as each receiving antenna of MIMO provides a signal measurement. The $l_0/l_1$ minimization method [9] is powerful, but the complexity is very high. To reduce the complexity, greedy iterative algorithm like block orthogonal matching pursuit (BOMP) [10] was proposed. To avoid matrix inversion in BOMP, approximate message passing MMV (AMP-MMV) [11] was proposed. Even when AMP-based methods are used, the complexity of large-scale matrix multiplications in all iterations is still very high. In our previous work [12], a non-iterative MMV solution named fast power reconstruction (FPR) was proposed, while the complexity is still much higher than grant-based one in massive MIMO. In a word, existing MMV solutions are too complex to implement in practice especially for mMTC using massive MIMO.

This letter first proposes a highly efficient pilot detection scheme using novel pilot detection reference signal (PDRS). Direct weight estimation (DWE) is also proposed to directly obtain spatial combining weight vectors without obtaining channel information. DWE is used to recover PDRS signals using potential pilots. Then, PDRS recovery errors are used for pilot detection. The performance of DWE is also theoretically analyzed and proved. Also, PDRS is only several-bit long and has little impact on the overall performance. The simulation results show that the proposed method gains an overall performance close to that of oracle detection using least squares (LS) channel estimation and zero-forcing (ZF) combining.

The proposed method is dedicated for mMTC using massive MIMO. In this scenario, the complexity becomes the bottleneck, while the reliability requirement is not strict. The major contributions of this letter include two aspects. First, the proposed PDRS-based method gains a great complexity reduction by orders of magnitude. It makes grant-free mMTC using massive MIMO becomes practical as the complexity is comparable to that of grant-based transmission. Second, the proposed DWE gains an invariant performance even when many inactive users are detected as active. That is to say, an aggressive detection of more users than the ground truth can be supported without performance degradation.

The rest of this letter is organized as follows. In Section II, the system model and existing works are introduced. In Section III, PDRS and DWE are proposed and analyzed. Section IV shows the numerical results to verify the performance and complexity. Throughout the letter, $(\cdot)^T$, $(\cdot)^H$ and $(\cdot)^+$ denote the transpose, Hermitian transpose, pseudo-inverse of a matrix, respectively. $\mathbf{I}_N$ denotes the unit matrix of size $N \times N$. $\mathbf{A}(\mathbf{x},:)$ and $\mathbf{A}(:,\mathbf{x})$ is the matrix consisting of the row vectors and column vectors with indices $\mathbf{x}$, respectively. $\|\cdot\|_2$ denotes $l_2$ norm. The

This study is supported by National Key R&D Program of China under Grants 2019YFB1803400.

Y. Ma, Z. Yuan, W. Li and Z. Li are with the Department of Wireless Algorithm, ZTE Corporation, Shenzhen, 518057, China; they are also with State Key Laboratory of Mobile Network and Mobile Multimedia Technology, Shenzhen, 518057, China (e-mail: yihua.ma@zte.com.cn; yuan.zhifeng@zte.com.cn; li.weimin6@zte.com.cn; li.zhigang4@zte.com.cn).

computational complexity is represented by the number of complex number multiplications.

## II. SYSTEM MODEL AND EXISTING WORKS

### A. System Model

Assume that the base station has $M$ antennas, while users have one. The non-orthogonal pilot pool is denoted by $\mathbf{P} = [\mathbf{p}_1^T, \mathbf{p}_2^T, ..., \mathbf{p}_N^T]^T \in \mathbb{C}^{N \times L}$ with pilot vector $\mathbf{p}_i \in \mathbb{C}^{1 \times L}$. $L$ and $N$ represent the length and number of pilots, and $L < N$. $N$ pilots in $\mathbf{P}$ are allocated to $N$ users, and the number of active users is $K$. The received pilot signal $\mathbf{Y} \in \mathbb{C}^{M \times L}$ is

$$\mathbf{Y} = \mathbf{H}\mathbf{P}_A + \mathbf{N} \quad (1)$$

where $\mathbf{H} = [\mathbf{h}_1, \mathbf{h}_2, ..., \mathbf{h}_N] \in \mathbb{C}^{M \times N}$, $\mathbf{h}_i \in \mathbb{C}^{M \times 1}$, $\mathbf{P}_A = [a_1\mathbf{p}_1^T, a_2\mathbf{p}_2^T, ..., a_N\mathbf{p}_N^T]^T \in \mathbb{C}^{N \times L}$, $a_i \in \{0, 1\}$ and $\mathbf{N} \in \mathbb{C}^{M \times L} \sim \mathcal{CN}(0, \sigma^2)$ denote the channel matrix, the channel vector of the $i$-th user, the active pilot matrix, the activity flag of the $i$-th user and complex Gaussian noise with a power of $\sigma^2$, respectively. When the $i$-th user is active, $a_i = 1$, and vice versa.

From the angle of CS, (1) becomes an MMV problem as

$$\mathbf{Y}^T = \mathbf{P}^T \mathbf{H}_A^T + \mathbf{N}^T \quad (2)$$

where $\mathbf{H}_A^T = [a_1\mathbf{h}_1, a_2\mathbf{h}_2, ..., a_N\mathbf{h}_N]^T \in \mathbb{C}^{N \times M}$. As user activity is sporadic, $\mathbf{H}_A^T$ contains only $K$ non-zero row vectors with $K < N$. With the sparsity feature, (2) can be solved via CS methods.

### B. Existing Methods

The joint user detection and channel estimation of (2) can be solved by iterative greedy methods. One representative method is BOMP [10], and it can be described by iterations as

$$\mathbf{s}^t = \mathbf{s}^{t-1} \cup \left( \arg\max_n \left\| \left(\mathbf{P}^T(n,:)\right)^H \left(\mathbf{Z}^{t-1}\right)^T \right\|_2 \text{ for } n \notin \mathbf{s}^{t-1} \right) \quad (3)$$

$$\left(\mathbf{Z}^t\right)^T = \mathbf{Y}^T - \mathbf{P}(\mathbf{s}^t,:)^T \left( \left(\mathbf{P}(\mathbf{s}^t,:)^T\right)^+ \mathbf{Y}^T \right) \quad (4)$$

where $\mathbf{s}^t$ is the support vector set containing the active user indices, and $\mathbf{Z}^t$ is the residual signal in the $t$-th iteration. The initial condition of iteration is that $\mathbf{Z}^0 = \mathbf{Y}$, and $\mathbf{s}^0$ is empty. The number of iterations in BOMP is fixed to $K$.

The matrix inversion in (4) in each iteration is complex, and AMP-MMV [11] was proposed to avoid the matrix inversion as well as achieve a nearly optimal solution. It can be briefly expressed by the following iterations as

$$\left(\mathbf{H}^t\right)^T = F\left(\mathbf{H}^{t-1} + \mathbf{P}\left(\mathbf{Z}^t\right)^T\right) \quad (5)$$

$$\left(\mathbf{Z}^{t+1}\right)^T = \mathbf{Y}^T - \mathbf{P}^T\left(\mathbf{H}^t\right)^T + \mathbf{O}^t \quad (6)$$

where $F(\cdot)$ is a conditional expectation function, $\mathbf{H}^t$ is the iterative channel estimation and $\mathbf{O}^t$ is the so-called Onsager reaction term in the $t$-th iteration.

For MMV problem with a small $M$, AMP-based method greatly reduces the complexity as the matrix inversion takes up the major part of total complexity. However, in grant-free massive MIMO, $M$ is very large, which makes the matrix inversion complexity no longer dominate. The complexity of matrix inversion in (4) is $O(K \cdot K^3)$, while that of matrix multiplication in (3) is $O(K \cdot NLM)$. In grant-free massive MIMO for mMTC, $N \gg K$, $M > K$, and $L \sim K$, which means $O(K \cdot NLM) \gg O(K \cdot K^3)$. $\sim$ denotes that two numbers are in the same order of magnitude. That is to say, the matrix inversion complexity reduction is not enough for massive MIMO. Both BOMP and AMP-MMV have a complexity of $O(N_{it} \cdot NLM)$, where $N_{it}$ is the total number of iterations.

Apart from iterative solvers, a non-iterative solver named FPR [12] was used to solve MMV problems. The first step of FPR is a matched filtering (MF) channel estimation as

$$\mathbf{H}_{MF} = \mathbf{Y}\mathbf{P}^H \quad (7)$$

The power of the MF channel estimation is $\mathbf{p}_{MF} = [p_{m1}, p_{m2}, ..., p_{mN}]$, where $p_{mi} = \|\mathbf{H}_{MF}(:,i)\|_2^2$. Then, the channel power of each potential pilot is recovered to decide the active pilot support as

$$\mathbf{s} = \{n_1, n_2, ..., n_\xi\} = \arg\max_n \mathbf{p}_R(n) \quad (8)$$

where $\mathbf{p}_R = \mathbf{p}_{MF}\left((\mathbf{P}\mathbf{P}^H) \circ (\mathbf{P}\mathbf{P}^H)^{\circ H}\right)^+$ is the recovered power vector, $\mathbf{p}_R(n)$ is the $n$-th element of vector $\mathbf{p}_R$, $\{n_1, ..., n_\xi\}$ is $\xi$ detected pilot indices, $\xi$ is the number of detected pilots, $\circ$ denotes the Hadamard product, and $(\cdot)^{\circ H}$ returns a matrix of element-wise conjugate. The pseudo-inverse of $\mathbf{P}_R$ is fixed as $\mathbf{P}$ is certain. Therefore, it can be calculated in advance and stored in hardware, which does not cause any computational complexity. FPR was used for spread code detection. In grant-free pilot detection, the pseudo-inverse result contain $L^2$ complex numbers, which leads to a high storage cost. Also, the complexity of FPR, $O(NML)$, is still much larger than that of grant-based ZF combination, $O(K^3)$. This letter aims to finding a solution with a computational complexity close to $O(K^3)$.

## III. THE PROPOSED PDRS SCHEME

### A. The proposed PDRS scheme

To solve the complexity problem, PDRS is proposed. As shown in Fig. 1(b), PDRS is an extra signal to assist pilot detection. The length of PDRS is assumed to be $l$. PDRS codebook design is very tolerant. It can be an orthogonal codebook, which means $K$ users select $l$ codes with repeated selections. It can also be randomly generated with normalized power. With loss of generality, PDRS matrix is denoted by $\mathbf{R} = [\mathbf{r}_1^T, \mathbf{r}_2^T, ..., \mathbf{r}_N^T]^T \in \mathbb{C}^{N \times l}$, and vectors in $\mathbf{R}$ can be repetitive. Then, the received signal consisting of both PDRS and pilot is

$$\mathbf{Y}' = [\mathbf{Y}_R, \mathbf{Y}] = \mathbf{H}[\mathbf{R}_A, \mathbf{P}_A] + \mathbf{N}' \quad (9)$$

where $\mathbf{Y}_R \in \mathbb{C}^{M \times l}$, $\mathbf{R}_A = [a_1\mathbf{r}_1^T, a_2\mathbf{r}_2^T, ..., a_N\mathbf{r}_N^T]^T \in \mathbb{C}^{N \times l}$, and $\mathbf{N}' \in$

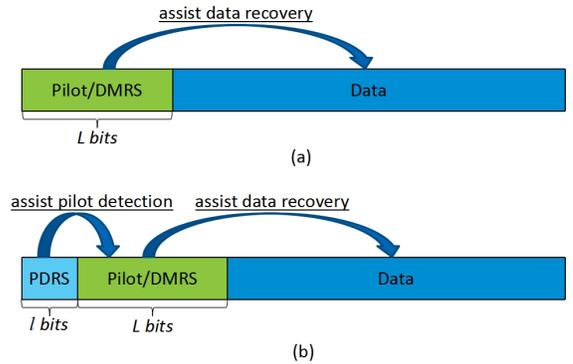

Fig. 1. The packet structure of (a) the conventional grant-free scheme, and (b) the proposed PDRS grant-free scheme

$\mathbb{C}^{M \times (L+l)} \sim \mathcal{CN}(0, \sigma^2)$ is complex Gaussian noise.

In conventional methods, spatial combining weights require to be calculated after pilot detection and channel estimation. To quickly recover signals, DWE is proposed. An LS estimation of spatial combining weights is directly applied as

$$\mathbf{W} = \mathbf{P}\mathbf{Y}^+ \quad (10)$$

where $\mathbf{W} = [\mathbf{w}_1^T, \mathbf{w}_2^T, ..., \mathbf{w}_N^T]^T \in \mathbb{C}^{N \times M}$, and $\mathbf{w}_i \in \mathbb{C}^{1 \times M}$ is the spatial combining weight vector of the $i$-th pilot. Using DWE, all $N$ pilots in $\mathbf{P}$ can have their own spatial combining weight vectors. The relationship between DWE and the conventional scheme will be analyzed in the next sub-section.

With estimated weights for all potential pilots, the pilot activity possibility is evaluated via recovery errors of PDRS. Then, the support vector is decided via one-shot sorting as

$$\mathbf{s} = \{n_1, n_2, ..., n_\xi\} = \arg\min_n \|\mathbf{w}_n\mathbf{Y}_R - \mathbf{r}_n\|_2^2 \quad (11)$$

In (11), channel hardening feature of massive MIMO reduces the impact of weight estimation errors on PDRS recovery. This feature works similarly for data recovery in general schemes. This method also works when an aggressive detection of $\xi > K$ is made to increase the true positive rate with the cost of increasing false positive detection. True positive detection is that an active pilot is detected to be active, while false positive detection is that an inactive pilot is detected to be active. The false positive detection may bring some invalid decoding complexity overheads, but the impact is acceptable due to two reasons: (a) the final performance is not affected as wrong streams cannot pass cyclic redundancy check (CRC), and (b) the decoding of wrong signal streams can be easily avoided using the data features [13] like post signal to interference and noise ratio (SINR) of a small fraction of data.

Combining (10) and (11), the major complexity of PDRS-based pilot detection is from calculating $\mathbf{P} \cdot \mathbf{Y}^+ \cdot \mathbf{Y}_R$. The second matrix multiplication in $\mathbf{P} \cdot \mathbf{Y}^+ \cdot \mathbf{Y}_R$ is calculated before the first one to reduce total complexity. In this way, the complexity of the proposed scheme is $O(NMl+L^3)$. Moreover, the proposed scheme is able to directly obtain combining weights, while other schemes require extra calculations. The whole processing procedure at the receiver side can be very simple, including 3 steps: (a) pilot detection using one-shot sorting of PDRS recovery errors, (b) spatial combination using DWE weights, and (c) demodulation of combined data. Step (b) can also be substituted by conventional combining methods.

*B. Analysis of Direct Weight Estimation*

To verify the performance of DWE, two conclusions are given and then proved.

**Lemma 1.** *When the number of detected users $\xi$ is not smaller than the pilot length $L$, DWE is equivalent to LS channel estimation plus ZF spatial combining.*

*Proof:*

Assume that the detected pilot matrix is $\mathbf{P}_{det} = [\mathbf{p}_{d1}^T, \mathbf{p}_{d2}^T, ..., \mathbf{p}_{d\xi}^T]^T \in \mathbb{C}^{\xi \times L}$, where $di$ is the index of detected pilot/user. LS channel estimation is

$$\hat{\mathbf{H}}_{LS} = \mathbf{Y}\mathbf{P}_{det}^+ \quad (12)$$

The corresponding ZF weight matrix is the pseudo-inverse of estimated channel as

$$\mathbf{W}_{ZF} = \hat{\mathbf{H}}_{LS}^+ = \left(\mathbf{Y}\mathbf{P}_{det}^+\right)^+ \quad (13)$$

Using pseudo-inverse property that $(\mathbf{AB})^+ = (\mathbf{A}^+\mathbf{AB})^+(\mathbf{ABB}^+)^+$ [14], (13) becomes

$$\mathbf{W}_{ZF} = \left(\mathbf{Y}^+\mathbf{Y}\mathbf{P}_{det}^+\right)^+\left(\mathbf{Y}\mathbf{P}_{det}^+\mathbf{P}_{det}\right)^+ \quad (14)$$

As $M > L$ and column vectors of $\mathbf{Y}$ are uncorrelated, the rank of $\mathbf{Y}$ is $r = L$. Using SVD decomposition of $\mathbf{Y} = \mathbf{U}_r\mathbf{D}_r\mathbf{V}_r^H$, we can have

$$\mathbf{Y}^+\mathbf{Y} = \left(\mathbf{V}_r\mathbf{D}_r^{-1}\mathbf{U}_r^H\right)\left(\mathbf{U}_r\mathbf{D}_r\mathbf{V}_r^H\right) = \mathbf{I}_r \quad (15)$$

In Lemma 1, $\xi \geq L$, and the column vectors of $\mathbf{P}_{det}$ are uncorrelated. Therefore, similarly to (15), $\mathbf{P}_{det}^+\mathbf{P}_{det} = \mathbf{I}_L$. Then, (14) is proved to be equivalent to (10) as

$$\mathbf{W}_{ZF} = \left(\mathbf{Y}^+\mathbf{Y}\mathbf{P}_{det}^+\right)^+\left(\mathbf{Y}\mathbf{P}_{det}^+\mathbf{P}_{det}\right)^+ = \mathbf{P}_{det}\mathbf{Y}^+ \quad (16)$$

Therefore, Lemma 1 is proved.

**Lemma 2.** *When the number of active users $K$ is smaller than the pilot length $L$, the multi-user interference suppression of DWE for true positive users can be equivalent to a combination of oracle detection, LS channel estimation and ZF filter.*

*Proof:*

In Lemma 2, only the multi-user interference suppression performance is considered. Therefore, a simplification of $\mathbf{Y} = \mathbf{H}\mathbf{P}_A$ is made in the right pseudo-inverse matrix of (14). If an oracle detection of $\mathbf{P}_{det} = \mathbf{P}_A$ is gained, (14) can be simplified using (15) and the property that $\mathbf{A}\mathbf{A}^+\mathbf{A} = \mathbf{A}$ [14] as

$$\mathbf{W}_{ZF} = \left(\mathbf{I}_r\mathbf{P}_A^+\right)^+\left(\mathbf{H}\mathbf{P}_A\mathbf{P}_A^+\mathbf{P}_A\right)^+ = \mathbf{P}_A\left(\mathbf{H}\mathbf{P}_A\right)^+ = \mathbf{P}_A\mathbf{Y}^+ \quad (17)$$

Compared with (10), the interference suppression performance of DWE for true positive users can be equivalent to a scheme using oracle detection, LS channel estimation and ZF filter. Therefore, Lemma 2 is proved.

DWE brings many advantages. The first is stability with false detection. Both false positive and false negative detection will not affect estimated weights of true positive detected pilots. The false negative is that an active pilot is detected to be inactive. The second is aggressive detection to improve the performance. As false positive detection does not affect the performance, the receiver can give an activity result of $\xi > K$ to improve the detection hit rate and overall performance. The last is that the calculation complexity is reduced as LS plus ZF contains two matrix pseudo-inverse, while DWE only requires one. The disadvantage of DWE is that the performance when $K < L$ is worse than that of conventional method using an oracle detection due to noise. However, it is not easy to gain oracle detection in practice as $K$ is unknown. Also, in mMTC, a large loading is more preferred which means $K$ should be large.

IV. NUMERICAL RESULTS

*A. Simulation Settings*

In simulations, the number of receiving antennas $M = 128$. The transmission packet carries 336-bit information. It takes up a bandwidth of 2 resource blocks in 1 subframe, i.e., 336 resource elements, and 2/7 of them are for pilots, i.e., pilot length $L = 96$. The number of pilots $N = 1000$. CRC length is 16 bits. Quadrature phase shift keying (QPSK) modulation and low density parity check (LDPC) channel coding are employed. That is to say, the code rates of existing works are 0.7, while that for PDRS scheme is slightly higher as extra $l$ bits are used. Flat fading channel is used as the packet is small. Assuming that a simple open-loop power control [3] is used to compensate large-scale fading, the expectations of receiving power are equal for different users [6]. The channels between users and receiving antennas have an identically independent Rayleigh distribution. Both non-orthogonal pilots and PDRS generate elements using an independent complex Gaussian distribution with their powers respectively normalized.

In practice, the number of active users $K$ is unknown, and an aggressive detection of $\xi > K$ should be used to ensure the performance. Wrong streams can be identified by data features like post SINR to prevent high invalid decoding complexity. The aggressive detection coefficient is denoted by $\alpha = \xi/K$. Existing works require $\alpha = 1$, and this setting is used here. It does not mean a known $K$ is required by PDRS-based methods. For example, the receiver can roughly estimate the distribution of user number and set an enough large $\xi$. LS+ZF denotes LS channel estimation plus ZF spatial combination. Miss detection rate is the probability of active users not being detected, and block error rate (BLER) is the probability of the transmission block, or packet, being successfully demodulated and decoded.

### B. Pilot Detection Performance

As this letter focuses on mMTC, a target BLER of under 0.1 is considered. For an oracle detection with LS+ZF, signal to noise ratio (SNR) = 4 dB satisfies this requirement with $K = 96$. To make detection errors not affect the overall performance, the miss detection rate of the proposed PDRS scheme should be much lower than BLER reference. The miss detection rate and the BLER reference are shown in Fig. 2(a), which is used to decide the value of $l$ for PDRS. To ensure that the miss detection rate is much smaller than BLER reference, 1/10 of BLER reference is also shown in Fig. 2(a). When the miss detection rate is lower than 1/10 of the BLER reference, BLER is seen as not being affected.

For $\alpha = 1$, PDRS requires $l \geq 4$ to achieve a miss detection rate lower than 1/10 of BLER reference, while for $\alpha = 2$, any $l$ can be used to fulfill this requirement. It can be more efficient to increase $\alpha$ instead of $l$ as wrong streams can be removed using data features, and a little invalid decoding is acceptable. Two settings of PDRS are decided: $l = 1, \alpha = 2$; and $l = 4, \alpha = 1$. For these two settings, the code rate increases from 0.7 to 0.703 and 0.712, respectively, which slightly affects the demodulation performance.

Fig. 2(b) shows the pilot detection performance comparison of existing works and the proposed schemes. Among them, AMP-MMV achieves the best performance, while BOMP gains the worst. BOMP easily gets wrong results as it suffers from large multi-user interference in the beginning iterations, which cannot be corrected in the following iterations. As mentioned before, the detection performance only requires to be much lower than the BLER reference, and the selected two settings meet this requirement in Fig. 2(b). The detection performance improvement via further increasing $l$ or $\alpha$ is not necessary as it cannot effectively improve overall performance.

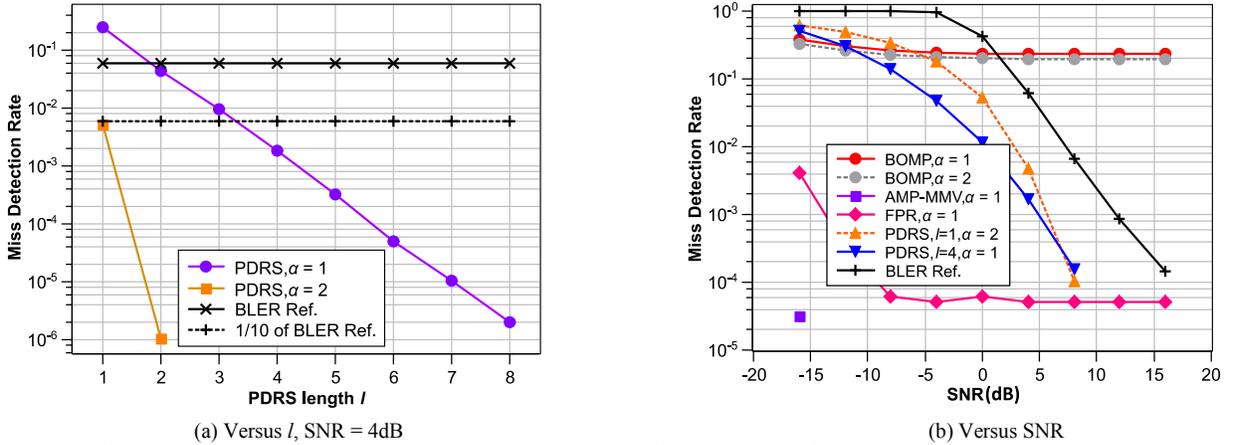

(a) Versus $l$, SNR = 4dB    (b) Versus SNR

Fig. 2. Miss detection rates of the proposed schemes and existing works. $K = 96$. The sum of miss detection rate and true positive rate is 1.

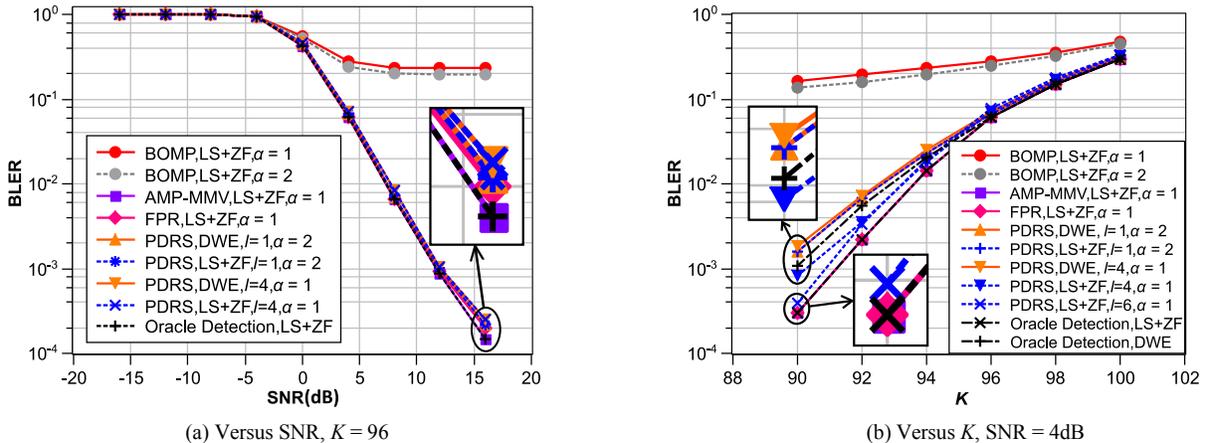

(a) Versus SNR, $K = 96$    (b) Versus $K$, SNR = 4dB

Fig. 3. BLER comparison of the proposed schemes and existing works. BLER of Oracle detection can be seen as a performance bound.

### C. Overall Performance

Fig. 3 shows BLER comparison of existing works and the proposed PDRS. In Fig. 3(a), the proposed PDRS schemes gain a BLER very close to that of the oracle detection. The BLER of PDRS is a little bit higher, which is caused by a slightly higher code rate. BLER of DWE equals to that of LS+ZF, which is proved in Lemma 1. AMP-MMV is able to achieve the BLER of the oracle detection, while BLER of FPR is slightly higher as the detection performance becomes comparable to the BLER reference when SNR is high in Fig. 2(b). In this simulation case, PDRS gains an overall performance very close to that of oracle detection using LS+ZF.

BLER comparison for different numbers of active user $K$ is shown in Fig. 3(b). PDRS provides a performance very close to that of the oracle detection when $K \geq L$. When $K < L$, BLER of oracle detection using DWE is higher than that of oracle detection using LS+ZF. It shows that the noise suppression of oracle detection using LS+ZF is better than that of DWE when $K < L$ as Lemma 2 has proven the equivalence of interference suppression. Therefore, PDRS schemes using DWE performs worse than LS+ZF bound. They are also slightly higher than DWE bound as the miss detection rate cannot be neglected when BLER decreases a lot with a smaller $K$. This degradation can be solved by replacing DWE with LS+ZF in data signal combining, which means DWE is only used in PDRS detection.

As shown in Fig. 3(b), PDRS using LS+ZF gains a better performance in the regions of $K < L$. PDRS using LS+ZF with $\alpha = 2$ gains the same BLER as that using DWE. Lemma 1 has shown this equivalence when $\xi \geq K$. PDRS using LS+ZF with $l = 4$ and $\alpha = 1$ outperforms the DWE bound, but it is still higher than LS+ZF bound due to miss detection errors. To eliminate the effect of miss detection, PDRS using LS+ZF with $l = 6$ and $\alpha = 1$ is added. With $l = 6$, the detection performance is good enough, and BLER becomes much closer to that of oracle detection using LS+ZF. Fig. 3 also shows that the proposed methods are more suitable for mMTC scenarios with ultra-high connection density and relatively low reliability requirement.

In practice, the ground truth of $K$ is not easy to gain, and the bound of oracle detection using LS+ZF will be hard to achieve. As a comparison, the bound of DWE is much easier to reach via an aggressive detection of $\xi > K$.

### D. Computational Complexity

A normalized complexity comparison is shown in Fig. 4. The complexity is normalized by the complexity of ZF, $O(K^3)$, which is the major complexity of grant-based massive MIMO. The number of iterations for AMP-MMV is obtained via simulations. The comparison shows that FPR reduces the computational complexity by 1~2 orders of magnitude compared with iterative methods including BOMP and AMP-MMV. However, it is still an order higher than that of grant-based transmission, and FPR has hardware storage problems as mentioned before. The proposed PDRS scheme further reduces the computational complexity by an order of magnitude. More importantly, it enables the complexity of grant-free massive MIMO to be comparable to that of grant-based one, which is crucial for practical implementations.

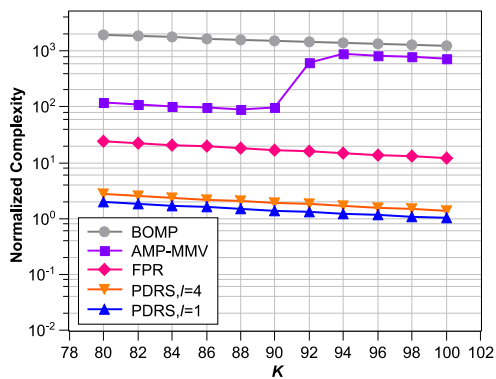

Fig. 4. Complexity comparison of the existing works and proposed schemes. The complexity is normalized by that of a grant-based transmission. $N_{iter}$ of AMP-MMV is obtained via simulations, which is set between 5 and 60.